\documentclass[12pt]{article}

\usepackage{graphicx}
\usepackage{caption}
\usepackage{times}

\newcommand{\bea}{\begin{eqnarray}}
\newcommand{\eea}{\end{eqnarray}}
\newcommand{\beq}{\begin{equation}}
\newcommand{\eeq}{\end{equation}}

\newenvironment{sciabstract}{%
\begin{quote} \bf}
{\end{quote}}

\newcounter{lastnote}

\title{Superconductivity-induced nematicity}

\author
{Y. S. Kushnirenko$^{1\ast}$,  D. V. Evtushinsky$^{1, 2}$, T. K. Kim$^{3}$,\\
I.V. Morozov$^{1, 4}$, L. Harnagea$^{1, 5}$, S. Wurmehl$^{1}$, S. Aswartham$^{1}$,\\ A.V. Chubukov$^{6}$, S. V. Borisenko$^{1}$
\\
\\
\normalsize{$^{1}$ IFW-Dresden, Helmholtzstr. 20, 01069 Dresden, Germany}\\
\normalsize{$^{2}$ Institute of Physics, Ecole Polytechnique Federale Lausanne, CH-1015 Lausanne, Switzerland}\\
\normalsize{$^{3}$ Diamond Light Source, Harwell Campus, Didcot OX11 0DE, United Kingdom}\\
\normalsize{$^{4}$ Lomonosov Moscow State University, 119991 Moscow, Russia}\\
\normalsize{$^{5}$ Indian Institute of Science Education and Research, Pune, Maharashtra-411008, India}\\
\normalsize{$^{6}$ Department of Physics, University of Minnesota, Minneapolis, USA}\\
}

\date{}

\begin{document} 
\baselineskip24pt

\maketitle

\begin{sciabstract}
The role of nematic order for the mechanism of high-temperature superconductivity is highly debated. In most iron-based superconductors (IBS) the tetragonal symmetry is broken already in the normal state, resulting in orthorhombic lattice distortions, static stripe magnetic order, or both. Superconductivity then emerges, at least at weak doping, already from the state with broken $C_4$ rotational symmetry. One of the few stoichiometric IBS, lithium iron arsenide, superconducts below 18 K and does not display either structural or magnetic transition in the normal state. Here we demonstrate, using angle-resolved photoemission, that even superconducting state in LiFeAs is also a nematic one. We observe spontaneous breaking of the rotational symmetry in the gap amplitude on all Fermi surfaces, as well as unidirectional distortion of the Fermi pockets. Remarkably, these deformations disappear above superconducting $T_c$. Our results demonstrate the realization of a novel phenomenon of superconductivity-induced nematicity in IBS, emphasizing the intimate relation between them. We suggest a theoretical explanation based on the emergence of a secondary instability inside the superconducting state,  which leads to the nematic order and s-d mixing in the gap function.
\end{sciabstract}

Several classes of materials,  which become superconducting at elevated temperatures, also show a spontaneous  unidirectional order in some part of their phase diagrams.  Examples range from stripes, directly observed in the cuprates \cite{tranquada1995evidence} and in FeSe films, \cite{li2017stripes} to nematic liquid state in the ruthenates \cite{borzi2007formation} and to rotational symmetry breaking state in iron-based superconductors (IBS) \cite{chu2010plane}. Nematicity has been one of the central topics in the studies of IBS during the last decade \cite{chubukovhirschfeld2015,fernandes2014drives}. It leads to significant anisotropy of the magnetic properties \cite{baek2015orbital} and of the electronic transport \cite{chu2010plane}, orthorhombic distortions of the lattice \cite{khasanov2010iron,rotter2008spin,huang2008neutron,goldman2008lattice}, and sizable changes in the low energy electron dynamics, e.g., band splitting close to the Fermi level, which  in  FeSe well exceeds the superconducting gap \cite{fedorov16, watson2016evidence}. Raman and other data \cite{blumberg2016critical, Gallais2016charge, yamakawa2016nematicity, benfatto2018nematic, fanfarillo2017nematicity, chinotti2017optical, toyoda2018nematic} on several  IBS reveal a strong increase of the nematic susceptibility, which starts well above the nematic transition temperature. The issue, which received less attention until recently, is the relation between nematicity and superconductivity.  From theory perspective, nematic fluctuations can mediate superconductivity \cite{lederer2015enhancement,klein2018dynamical}, and long range nematic order affects both the gap structure and superconducting $T_c$, as recent studies of FeSe$_{1-x}$S$_x$ have demonstrated \cite{Shibauchi2017maximizing}. In this work we discuss whether superconductivity can in turn induce a nematic order.

A natural candidate to address this question is stoichiometric lithium iron arsenide (LiFeAs). It is tetragonal in the normal state, and shows no magnetic or structural transition before it becomes superconducting at T$_c$=18 K \cite{borisenko2010superconductivity}. Earlier data in the superconducting state were interpreted assuming that $C_4$ symmetry remains intact.  It has been  shown recently that application of strain induces rotational symmetry breaking in the superconducting state of LiFeAs \cite{yim2018discovery}. We show here that in the superconducting state LiFeAs actually develops a spontaneous nematic order. 
   
\section*{Results}
In order to study possible signs of nematicity in superconducting LiFeAs we revisit its electronic structure and especially the gap function, using angle-resolved photoemission (ARPES) with a new level of precision.

In {\bf Fig. 1A} we show a Fermi surface (FS) map which roughly covers the 1-Fe Brillouin zone, often used in theoretical studies. This map represents all main features of the electronic structure of LiFeAs. The "dumbbell" in the center and the corresponding four-points feature in the corner at approximately  (-1.2, -1.2) are associated with the small $d_{xz}/d_{yz}$ hole-like pocket. The large square with rounded corners, also centered at $\Gamma$-point, is the $d_{xy}$ hole-like Fermi surface, and the pockets at the top and at the bottom of  {\bf Fig. 1A} are electron-like Fermi surfaces, formed by the $d_{xy}$ orbital and either $d_{xz}$ or $d_{yz}$ orbital. We will refer to the coordinate system of {\bf Fig. 1A} throughout the paper. Panels of {\bf Fig. 1B} show the temperature evolution of the characteristic high-symmetry cut, indicated on the map ({\bf Fig. 1A}) by the dashed orange line, which runs through all four Fermi surface sheets. From left to right, the dispersions correspond to $d_{xy}$-inner electron pocket, $d_{yz}$-outer electron pocket, $d_{xy}$-large hole pocket, $d_{xz}$- small hole pocket, and $d_{yz}$-dispersion, which does not cross the Fermi level. Each of the two latter dispersions  changes its orbital character between $d_{xz}$ and $d_{yz}$ under a rotation in  XY plane, but  has a particular orbital character  ( $d_{xz}$ or $d_{yz}$ ) along  high-symmetry directions. Therefore we label these dispersions and corresponding FS pockets as $d_{xz}$ or $d_{yz}$. As expected, the gap opens up at 17 K and gradually whips out the spectral weight from the Fermi level as the temperature is lowered. It is seen from the presented data that the largest superconducting gap is on the small hole-like pocket. It is about 5.4 meV at this particular $k_z$, as measured by fitting the corresponding energy-distribution curve [see supplementary materials \cite{SM_this} section II]. The next in magnitude is the gap on the inner electron pocket ($\sim$ 3.6 meV), and the smallest one is on the large $d_{xy}$-Fermi pocket ($\sim$ 2.3 meV). Because this  large hole pocket shows up in ARPES as a single dispersion, well separated from other dispersing features, the characteristic bending back of the dispersion is clearly seen in the lowest panel of {\bf Fig. 1B}. To underline the precision of our measurements, we zoom-in to this minimal gap and show the result in {\bf Fig. 1C} together with two typical energy-distribution curves (EDC) from the k-points marked on the map by red and magenta circles. Not only the sharpness of the EDCs, but also the presence of the coherence peaks above the Fermi level ( {\bf Fig. 1, B and C}), demonstrate that the superconducting gap in LiFeAs can be measured by ARPES with a very high precision (for details of the gap extraction from the data see \cite{SM_this} section II).

First, we consider in detail the features  associated with the hole pockets at the center of the BZ. The high-resolution dataset, shown in {\bf Fig. 2A}, is recorded under special geometry conditions to minimize the influence of the matrix element effects. If one compares the maps from {\bf Fig. 1A} and from {\bf Fig. 2A}, one can immediately notice that the pronounced minima along $k_x$ and $k_y$ in the former are absent in the latter. This is achieved by rotating the sample by 22.5$^{\circ}$. In this geometry the $d_{xy}$-states are not strongly suppressed along any direction in the k-space, providing a suitable non-symmetrized dataset for the gap extraction from the EDC-lineshape. The intensity in the map is still slightly asymmetric, but this has no influence on the lineshape of EDC. The gaps extracted from these EDCs are plotted in {\bf Fig. 2C} as  functions of the angle along the Fermi surfaces.

One of the central results of the present paper is immediately seen from this graph: the gap function does not obey the $C_4$ symmetry and has only two maxima and two minima signalling the $C_2$ rotational symmetry breaking. We emphasize that the amplitude of the gap oscillations is considerable, well above the error bars. The gap modulation cannot be described by a single cosine function indicating the presence of higher harmonics. Another observation, overlooked in the earlier studies, is the deformation of the Fermi surface itself. In {\bf Fig. 2B} we show the intensity distributions along $k_x$ and $k_y$ cuts (where $k_x$ and $k_y$ correspond to the coordinate system introduced in  {\bf Fig. 1A}). Momentum distribution curves from the Fermi level ($E_F$-MDC) clearly indicate that the large hole-like Fermi surface is elongated in the $k_y$-direction. Moreover, this conclusion is supported not only by the MDCs from the Fermi level. In {\bf Fig. 2D} we plot the  position of the maxima of MDC as a function of the binding energy.  The plot demonstrates that the distortion persists to higher binding energies. We define a deformation coefficient as $D= 2\frac{L_y-L_x}{L_y+L_x}$, where $L_y$ and $L_x$ are the pocket sizes (2k$_F$) in X and Y directions, respectively. Its average value for the binding energy interval shown in the inset, is 4 \%. 

In {\bf Fig. 2, E and G} we  show the results for the two $d_{xz}/d_{yz}$ hole dispersions near $\Gamma$. The results are similar to the ones for the $d_{xy}$ pocket, but there are important differences. Strictly speaking, neither of the $d_{xz}/d_{yz}$ dispersions cross the Fermi level at this $k_z$, which is near $\Gamma$-point [see \cite{SM_this} section III]. Still, one of the dispersions comes close enough to the Fermi level and thus "feels" the gap. The gap function extracted from the change of this dispersion below $T_c$ is in anti-phase with the one for $d_{xy}$-Fermi pocket ({\bf Fig. 2E}), and  it also has a clear $C_2$ symmetry rather than $C_4$. Since the tops of both $d_{xz}/d_{yz}$  dispersions are close to the Fermi level, the extraction of the distortions from the  MDC dispersion near the Fermi level is quite complicated, and  we have estimated the distortions by analyzing them at higher binding energies ({\bf Fig. 2G}). The distortions of the two $d_{xz}/d_{yz}$ dispersions turned out to be of opposite sign and reached 7.0\% and -2.4 \%. Since for the steeper dispersion the distance between the maxima of MDCs is smaller and their widths are larger, the error bars are larger.

Where does the observed distortions come from? {\bf Fig. 2F} demonstrates the temperature evolution of the dispersion upon crossing $T_c$. The evolution is highly atypical for a superconductor. Usually \cite{norman2001momentum,chubukov2004dispersion}, the dispersion in the superconducting state develops a stronger kink at higher binding energies and then runs vertically within the gap region and hits the Fermi level exactly at $k_F$, representing the so-called S-shaped dispersion. The data in {\bf Fig. 2F}, taken along $k_y$,  show no S-shape, but the the size of the Fermi surface grows in this particular direction. This effect is absent in the data taken above $T_c$, see {\bf Fig. 2H}, which implies that the deformation is caused by superconductivity.

To have a complete overview of the rotational symmetry breaking in LiFeAs, we extended our high-resolution measurements to electron-like pockets. We show in {\bf Fig. 3A} the FS map, taken using 25 eV photons, which corresponds to $k_z$s closer to the $\Gamma$MX-plane of the BZ [see \cite{SM_this} section III]. As it was found earlier \cite{Borisenko16NPh}, because of the spin-orbit interaction, the electron pockets hybridize along the lines, which connect them, and therefore are better described as inner and outer pockets rather than as crossed ellipses. For all $k_z$ values, the inner pocket in LiFeAs is of $d_{xy}$-character and the outer one is of $d_{xz,yz}$-character. This is because the crossing of the bands, coming from the bottoms of electron pockets, is below E$_F$ in the $\Gamma$MX-plane  \cite{Borisenko16NPh}. In spite of the increased $k_z$-resolution, the outer electron pocket in LiFeAs still appears blurred on the maps, where the $k_z$-dispersion is strong. Because of this, the gap function, shown in {\bf Fig. 3C}, contains more datapoints for the inner pocket then for the outer one. Nevertheless, both gaps are again two-fold symmetric with strong modulation amplitude. The degree of the gap variations is easy to see directly from the EDCs in the inset. These EDC's are taken from the two k-points marked on the map by small crosses. Again, the E$_F$-MDCs ({\bf Fig. 3E}) show that the inner pocket is deformed and is longer along $k_y$.  To analyze the outer electron pocket, we used incident photon energies  h$\nu$ = 23 eV and 21 eV.  At h$\nu$ = 23 eV the outer electron pocket is larger and better distinguishable from the inner one ({\bf Fig. 3B}). Fig  ({\bf Fig. 3D}) clearly shows that the gap on this pocket is two-fold symmetric. The E$_F$-MDCs ({\bf Fig. 3F}) show that the inner pocket is again elongated. Underlying dispersions ({\bf Fig. 3G}) yield the average distortion coefficient $D =$ 3.6 \%. 

At h$\nu$ = 21 eV  the outer pocket is even larger and better  separated from the inner one ({\bf Fig. 3H}). Two cuts along $k_x$ and $k_y$ show the underlying dispersions ({\bf Fig. 3I}) and it is seen that, at least for the $k_x$ cut, the dispersion features corresponding to the outer FS are much better defined.  Panels ({\bf Fig. 3K}) and ({\bf Fig. 3L}) show the temperature dependence of the dispersions and one can now notice that the size of the outer pocket along $k_x$ becomes noticeably larger upon entering the superconducting state. Nearly no change occurs between above and below $T_c$ along the $k_y$ direction. Interestingly, the distortion of the inner $d_{xy}$-pocket is now different ( {\bf Fig. 3J}) -- the distortion coefficient D becomes negative. 

Comparing the data from {\bf Fig. 3, K and L} with the ones presented in {\bf Fig. 2F} one can notice the drastically different temperature evolution of the dispersion at different places in the k-space. Superconductivity can bend it forward, back, or leave it practically untouched. Remarkably, the kinks in the dispersion \cite{kordyuk2011angle} are most pronounced where the gap is the largest.

We summarize our experimental observations in {\bf Fig. 4}, where we show all gap anisotropies and FS distortions. A sketch of the Fermi surface of LiFeAs in the normal state is given in {\bf Fig. 4A} together with the orbital composition of the pockets. There is no small $d_{xz}/d_{yz}$-pocket at $\Gamma$ because $d_{xz}/d_{yz}$ hole dispersions only approach the Fermi level without crossing it. The inner (outer) electron pockets are formed by $d_{xy}$ ($d_{xz, yz}$) orbitals at all $k_z$s'.  In {\bf Fig. 4B} the observed gap variations are shown as the thickness of the Fermi contours. The minimal thickness corresponds to the minimal gap. Distortions are shown schematically, qualitatively reproducing the behaviour of the deformation D. When the pocket size along  $k_y$ is larger, D is positive, when  it is larger along $k_x$, D is negative. Question marks indicate that the distortion of the outer electron pocket is somewhat difficult to determine because of the broadening caused by strong $k_z$-dispersion. Different signs of the distrortion of electron pockets at $k_z=0$ and $k_z$ = $\pi$ (different directions of arrows in {\bf Fig. 4B}) may be due to the existence of the additional in-plane interaction channel at $k_z = \pi$ because at this $k_z$ hole $d_{xz}/d_{yz}$ dispersions cross the Fermi level. While the detailed $k_z$-dependence of the observed effects still needs to be refined, calling for further, even more thorough experimental studies, {\bf Fig. 4} provides an overview of a spontaneous rotational symmetry breaking in the superconducting state of LiFeAs.

\section*{Discussion}
We now present theoretical analysis of the observed variation of the gap on hole and electron pockets.  The experimental facts most relevant to the analysis below are (i) the absence of the gap nodes on the $d_{xy}$ hole pocket, and (ii) the $\cos {2\theta}$ variation of the gap along this pocket. The first observation implies that the gap is not a pure d-wave, the second indicates that a  d-wave gap component is present along with an s-wave component, i.e., $\Delta_{xy}  (\theta) = \Delta_s + \Delta_d \cos{2\theta}$.  Such behavior is indeed expected when the system has a nematic order. Indeed,  once $C_4$ symmetry is broken, $s-$wave and $d-$wave gap components are no longer orthogonal, and the Landau Free energy in general contains the symmetry-allowed term $\Delta_s \Delta_d$, linear in both s-wave and d-wave gap components. Because of bilinear coupling, once one pairing component develops, it acts as a field for the other component, and, as a result, both are present.

The gap structure in LiFeAs has been analyzed in several papers \cite{ahn2014superconductivity, yin2014spin, saito2014reproduction, wang2013superconducting}.  Like we said, the electronic structure of this material is somewhat different from those of other Fe-pnictides in that  cylindrical pockets in  LiFeAs, which exist for all $k_z$ values, are the two electron pockets and the $d_{xy}$  hole pocket, centered at $k_x=k_y = \pm \pi$ in 1-Fe Brillouin zone,  while $d_{xz}/d_{yz}$ hole pockets, centered at $k_x=k_y=0$   exist only around $k_z = \pi$.  This electronic structure allows a competition between a number of possible pairing states, ranging from a conventional $s^{+-}$ with sign change between all hole and all electron pockets, to orbitally antiphase $s^{+-}$  with sign change between $d_{xz}/d_{yz}$ and $d_{xy}$ hole pockets (and additional sign change  for the gap on a hole pocket and a portion of an electron pocket with the same orbital content), to several $d-$wave gap structures. Previous ARPES experiments were fitted better by an s-wave gap (the best fit is for type A $s^{+-}$ state in Ref. \cite{ahn2014superconductivity}), and  we assume that in the absence of nematicity the gap would be an s-wave. ARPES data reported here show that at $T = 23K$, slightly above $T_c =18K$, the system remains in the tetragonal phase, while the data taken at $7K$ inside the superconducting state show a nematic order.  Assuming that the tetragonal symmetry is not broken above $T_c$, we are left with two options -- either it gets broken at $T_c$, or at some $T < T_c$.  In both cases, s-wave superconductivity triggers $C_4$ symmetry breaking and the apperance of the $d-$wave component of the pairing gap. We didn't find a theoretical justification for the first scenario, but we did find the argument for the second one. 
   
Our analysis is similar to the one put forward by Fernandes and Millis \cite{fernandes2013nematicity}, but we employ somewhat different rational and go beyond their analysis in the computation of the parameters in the Free energy ${\cal F}$.  

Consider for definiteness the hole $d_{xy}$ pocket. Let us introduce a nematic order parameter $\Delta_n \cos {2 \theta}$.  Because $d-$wave gap component also scales as $\cos {2\theta}$, 
${\cal F}$  should generally contain the term
\beq
    \gamma \Delta_n \left(\Delta_s \Delta^*_d + \Delta^*_s \Delta_d \right)
    \label{s_1}
 \eeq
Let's suppose that an $s-$wave order develops on its own at $T_c$, while nematic order and d-wave superconducting order do not develop in the absence of $\Delta_s$.   The Free energy slightly below $T_c$ is then
\bea
 {\cal F} &=& \alpha_s |\Delta_s|^2 + \beta_s |\Delta_s|^4 +   \alpha_d  |\Delta_d|^2 + \beta_d |\Delta_d|^4 \nonumber \\
 && + \alpha_n |\Delta_n|^2 + \beta_n |\Delta_n|^4 +  2\gamma \Delta_n |\Delta_s| |\Delta_d| \cos {\phi} + ...
 \label{s_2}
\eea
where $\phi$ is the relative phase between $\Delta_s$ and $\Delta_d$, and dots stand for the terms which we will not need.  By construction, $\alpha_s <0$, while  $\alpha_{d,n} >0$ and $\beta_{s,d,n} >0$.
At $\gamma =0$, $|\Delta_s|^2 = - \alpha_s/(2\beta_s)$ and $\Delta_d = \Delta_n =0$. At a finite $\gamma$ (of either sign), the minimization with respect to $\Delta_s, \Delta_d, \Delta_n$, and $\phi$ yields
\bea
  -\alpha_s |\Delta_s| + |\gamma| |\Delta_n| |\Delta_d| &=&    2 \beta_s  |\Delta_s|^3 \nonumber \\
  -\alpha_s |\Delta_d|  + |\gamma| |\Delta_n| |\Delta_s|  &=&   2 \beta_d  |\Delta_d|^3 \nonumber \\ 
  -\alpha_n |\Delta_n|  + |\gamma| |\Delta_s| |\Delta_d|  &=&   2 \beta_n  |\Delta_n|^3 
 \label{s_3}
\eea 
At small negative $\alpha_s$ the solution is an $s-$wave order ($\Delta_d = \Delta_n =0$). However, as $|\alpha_s|$ increases, the system may simultaneously develop two other orders. This happens when 
\beq
  |\alpha_s|^{1/2} > \frac{(4 \beta_s |\alpha_n \alpha_d|)^{1/2}}{\gamma}, or \frac{|\alpha_s|}{|\alpha_n \alpha_d|^{1/2}} > \frac{4 \beta_s |\Delta_s|}{\gamma}
  \label{s_4}
\eeq
The inequality in Eq. (\ref{s_4})  is definitely satisfied below some $T < T_c$ if the tendency towards nematic order and/or d-wave superconducting order is strong, i.e., the product $|\alpha_n \alpha_d|$ is small. 
  
The coupling $\gamma$ is graphically represented as a triangular diagram with $\Delta_n, \Delta_s$, and $\Delta_d$ in the vertices and three internal fermionic lines with momenta/frequency  $(k, \omega), (k, \omega)$ and $(-k, - \omega)$. Evaluating the convolution of the three Green's functions with these momenta and frequency and assuming parabolic dispersion for fermions near the $d_{xy}$ hole pocket with $\epsilon_k = \mu - k^2/(2m)$, we obtain $|\gamma| = m/(16 \pi \mu) = 1/(8\pi v^2_F)$.  The coefficient $\beta_s$ is obtained in a similar manner by evaluating the square diagram with $\Delta_s$ in the vertices and four fermionic lines, two with $(k, \omega)$ and two with $(-k, -\omega)$. Evaluating the convolution of the four fermionic Green's functions in the same way as in \cite{fernandes2012preemptive} 
, we obtain $\beta_s = 7 m \zeta (3)/(16 \pi^3 T^2)$.  Substituting the expressions for $|\gamma|$ and $\beta_s$ into (\ref{s_4}), we obtain the condition for $s-$wave induced nematicity as
\beq
 \frac{|\alpha_s|}{|\alpha_n \alpha_d|^{1/2}} > \frac{28 \zeta(3)}{\pi^2}  \frac{\mu |\Delta_s|}{T^2}
 \label{s_5}
\eeq
For  $T \sim T_c \sim \Delta_s$ it becomes $|\alpha_s| > A |\alpha_n \alpha_d|^{1/2} (\mu/T_c)$, where $A \geq 1$. For a system in which $\mu/T_c$ is large,  the tendency towards nematic and/or d-wave instability near $T_c$ must be strong, otherwise the inequality on $|\alpha_s|$ would not be satisfied. In LiFeAs, however, all pockets are small and the probability that s-wave superconducting order will generate nematicity are much stronger.

Previous ARPES and STM studies of the gap anisotropy \cite{BorisenkoSymm,allan2012anisotropic,umezawa2012unconventional} in LiFeAs were interpreted as evidence for a pure $s-$wave gap with $\cos{4\theta}$ variation along the hole pockets. The presence of domains could be one possible explanation because of a finite spot size in ARPES and because a large enough area is needed to obtain a QPI pattern in STM. Another explanation could be poorer quality of earlier ARPES data. Finally, some of earlier data were actually obtained by using $C_4$ symmetrization procedure.

The observed spontaneous rotational symmetry breaking of the superconducting gap amplitude in LiFeAs is different from the symmetry breaking in d-wave or chiral-p-wave superconductors. In the latter only the phase acquires a new symmetry and there is no change in the macroscopic state of the system under rotation, therefore the rotational symmetry breaking can be detected only in interference experiments. In the present case the macroscopic state of the system does change by the rotation and thus such a symmetry breaking should be seen in bulk properties. Our data call for further, more detailed and phase-sensitive experiments on LiFeAs and other IBS.

\section*{Acknowledgements:}
We are grateful to Rafael Fernandes, Guenter Behr and Christian Hess for the fruitful discussions. We acknowledge Diamond Light Source for time on Beamline I05 under proposals NT5008 \& SI9689. YSK and SVB are supported by DFG Grants No. BO1912/6-1 and SPP1458. AVC is supported by the Office of Basic Energy Sciences, U.S. Department of Energy, under the award DE-SC0014402.

\clearpage

\begin{figure}[b]
	\centering
    \includegraphics[width=0.90\linewidth]{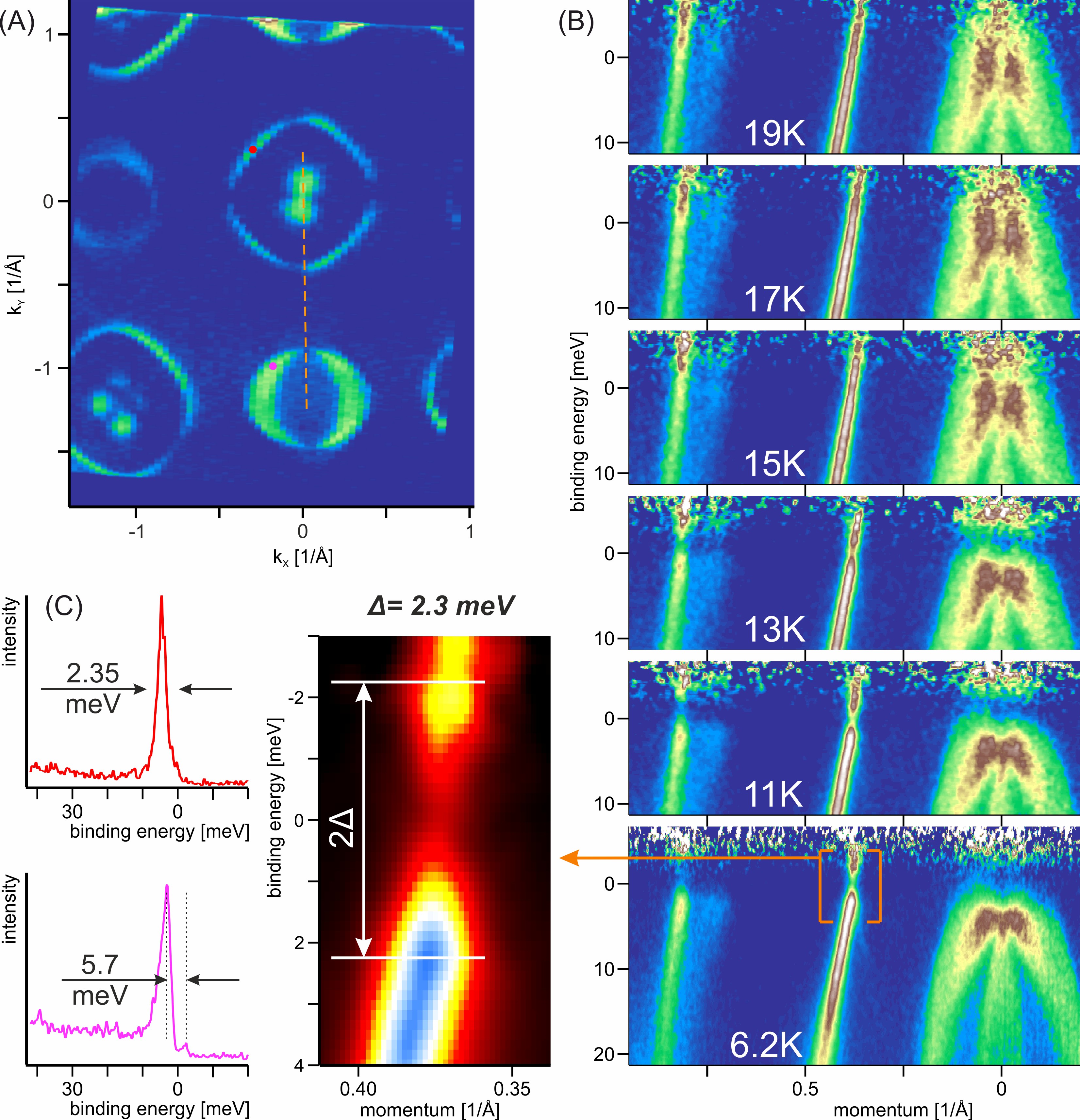}
	\caption*{{\bf Fig. 1} Superconducting gaps from ARPES. {\bf (A)} Overview Fermi surface map taken using 80 eV photons. {\bf (B)} Intensity plots corresponding to the dashed line in (A) measured as a function of temperature with 21 eV photons. All spectra are divided by the Fermi function to enhance the signal above the Fermi level. {\bf (C)} Exemplary EDCs from the k-points marked in (A) by red dots. 2D intensity plot corresponds to the area limited by the square brackets in the lowest panel of (B).}
	\label{fig:1}
\end{figure}

\clearpage

\begin{figure}[b]
	\centering
    \includegraphics[width=0.70\linewidth]{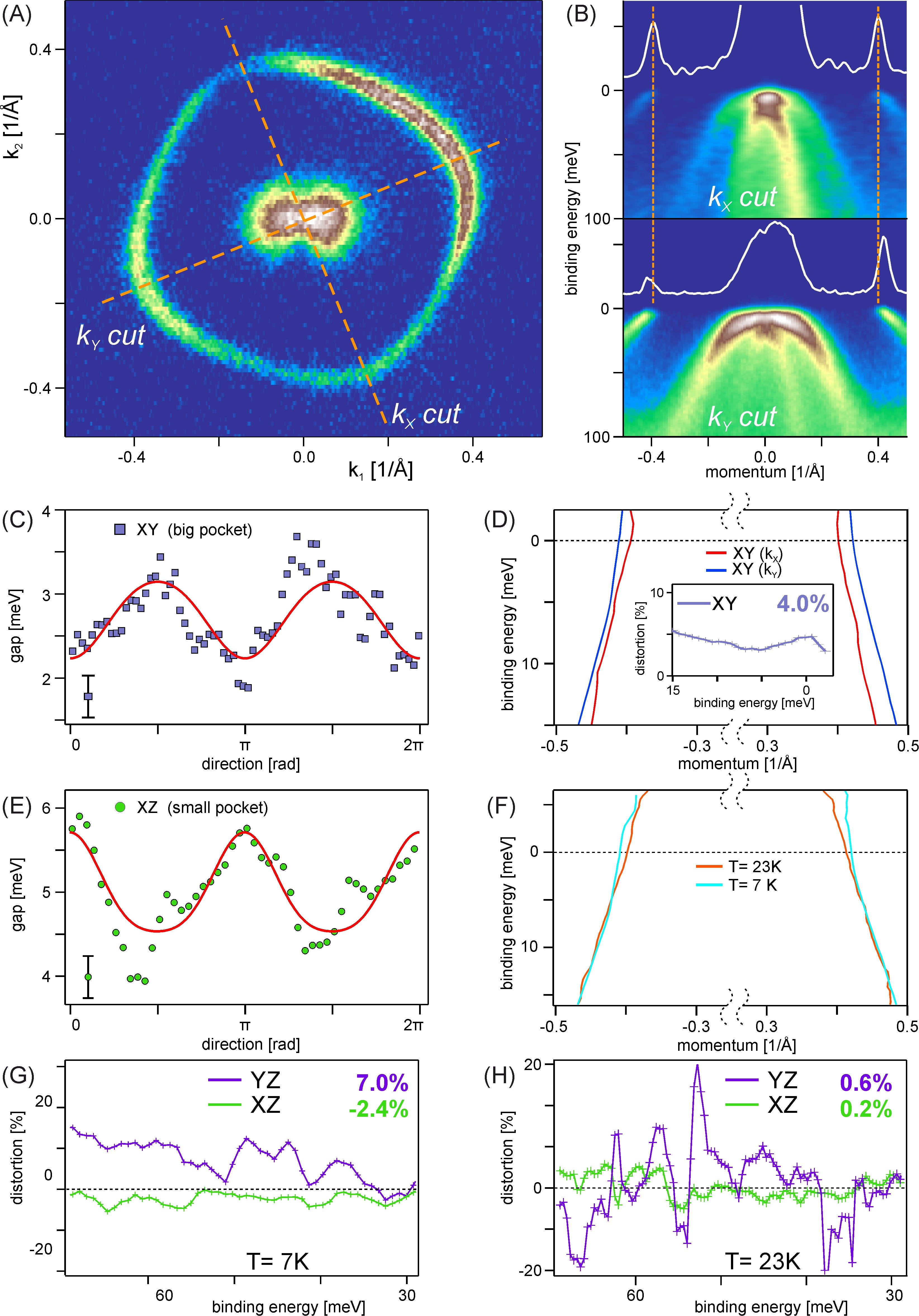}
	\caption*{{\bf Fig. 2} Gap anisotropy and distortions of the hole-like FS pockets and dispersions. {\bf(A)} High-resolution FS map measured at 7K with 25 eV photons. {\bf (B)} Intensity distributions along the $k_x$ and $k_y$ cuts. White curves are $E_F$-MDCs. Vertical dashed lines help to compare the peak positions. {\bf (C)} Gap function of the large hole-pocket. Here angle is counted anticlockwise from k$_y$ direction. Fitting function is mostly $\cos {2\theta}$ with a small ($\sim$ 6 \%) admixture of higher harmonics. {\bf (D)} Dispersions corresponding to large hole pocket extracted from (B). Inset shows the behavior of the distortion coefficient. {\bf (E)} Gap function of the small hole pocket. {\bf (F)} Temperature dependence of the dispersions corresponding to $k_y$-cut. {\bf (G)} Distortion coefficient for the xz and yz dispersions in the center of the BZ. {\bf (H)} The same as in (G), but measured at 23 K.}
	\label{fig:2}
\end{figure}

\clearpage

\begin{figure}[b]
	\centering
    \includegraphics[width=0.90\linewidth]{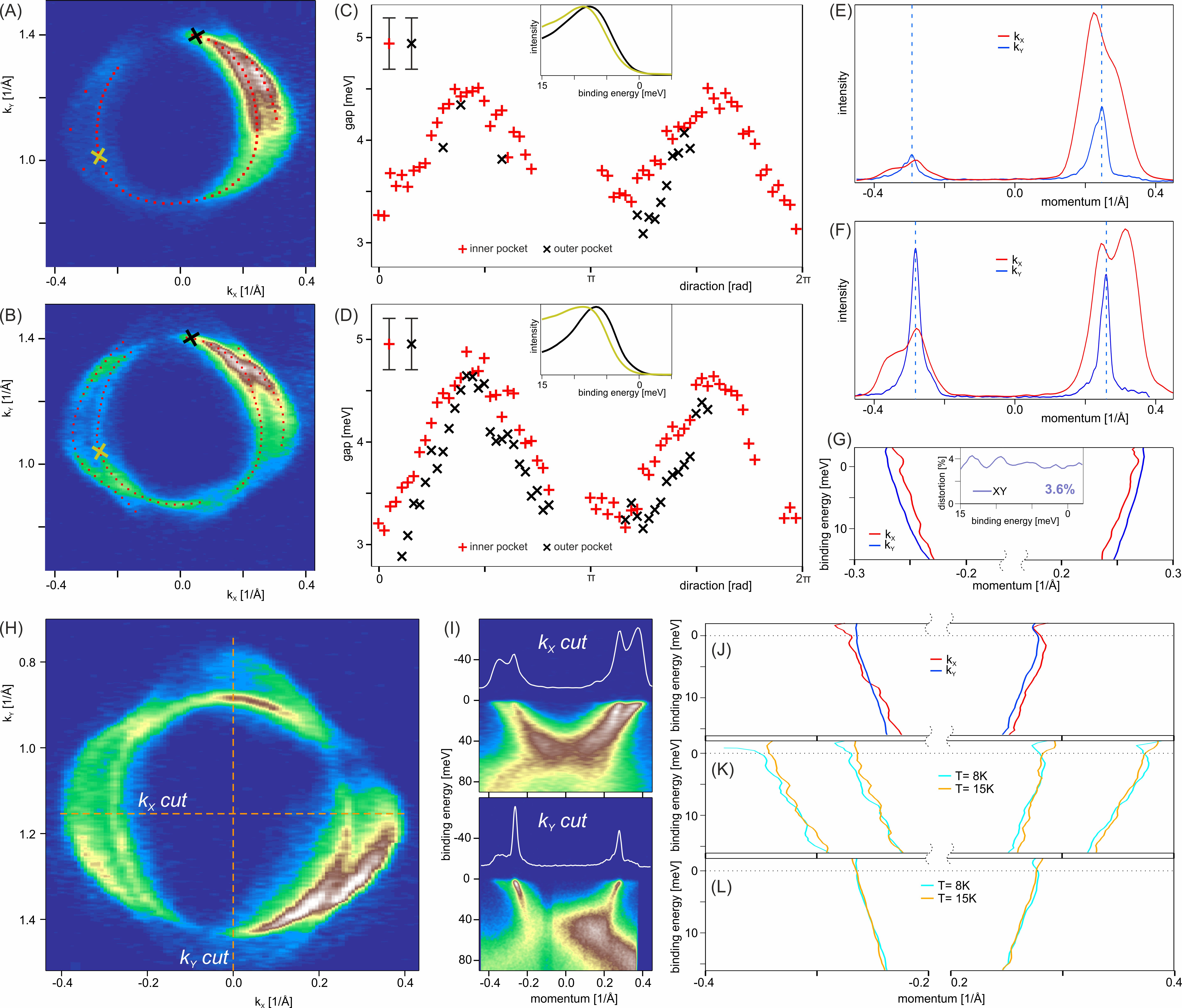}
	\caption*{{\bf Fig. 3} Gap anisotropy and distortions of the electron-like FS pockets. {\bf (A and B)} FS map of electron pockets measured with 25 eV and 23 eV photons respectively. {\bf (C and D)} Corresponding to (A) and (B) gap functions. Here angle is counted anticlockwise from k$_y$ direction. Insets show EDCs from the points on the maps marked by the crosses of the same color. {\bf (E and F)} Corresponding to (A) and (B) EF-MDCs. Dashed lines indicate the different positions of the peaks. {\bf (G)} Dispersions supporting the inner electron pocket from (B). Inset shows the distortion coefficient and its average value. {\bf (H)} FS map at 21 eV. (I) Intensity distribution along $k_x$- and $k_z$-cuts from (H) together with the corresponding EF-MDCs. {\bf (J)} Dispersions corresponding to the inner electron-pocket from (H). {\bf (K)} Temperature dependence of the dispersions along the $k_x$-cut. {\bf (L)} Temperature dependence of the dispersions of inner pocket along the $k_y$-cut. No matching of the zero position has been done in (J-L)}
	\label{fig:3}
\end{figure}

\clearpage

\begin{figure}[b]
	\centering
    \includegraphics[width=0.90\linewidth]{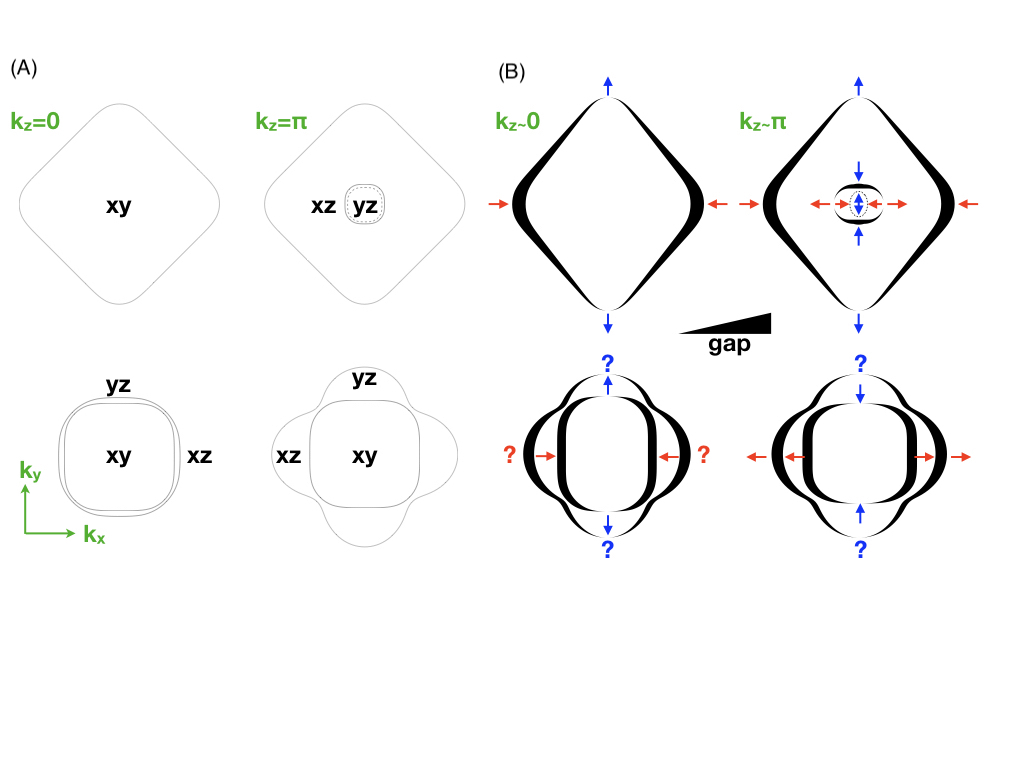}
	\caption*{{\bf Fig. 4} Nematic order in the superconducting state. {\bf (A)} Schematic Fermi surface contours of LiFeAs from the experiments in the normal state. Dashed contour represents the yz-states which do not cross the Fermi level. {\bf (B)} Qualitative sketch of the distortions and gap anisotropies consistent with the experimental data. Red (blue) arrows indicate squeezing (stretching) of the FSs. Question marks indicate uncertainty as regards the distortion of the outer electron pocket.}
	\label{fig:4}
\end{figure}

\clearpage

\section*{Supplementary Materials}
\subsection*{I. Materials and Methods}
LiFeAs single crystals in the form of packets of plates with dimensions of up to 1 cm were grown by self-flux using the standard method \cite{morozov2010single}. For the ARPES study single-crystal plates with dimensions of 3x3x0.1xmm3 have been selected. The preparation of single crystals for the measurement by the ARPES method was carried out in a dry argon box. Experiments have been carried out at I05 beamline of Diamond Light Source \cite{hoesch17RSI}. Single-crystal samples were cleaved \textit{in situ} in a vacuum better than $2\times10^{-10}$ mbar. Measurements were performed using linearly polarized synchrotron light, utilizing Scienta R4000 hemispherical electron energy analyzer with an angular resolution of 0.2$^\circ$ – 0.5$^\circ$ and an energy resolution of 2 meV. None of the maps presented in the paper are symmetrized.

\clearpage

\subsection*{II. Details of gap extraction}
EDCs in {\bf Fig. 1A} of the main text exhibit very narrow coherent peaks. In the magenta EDC one can clearly distinguish the second coherent peak, which is located above the Fermi level. The distance between the peaks is 5.7 meV, but this distance is smaller then the real doubled gap size because the shape of the second peak is heavily distorted by the Fermi function. In order to extract the real gap size one should fit this EDC with a function, which includes the influence of the Fermi function.

We fit EDC with a function which consists of 2 peaks multiplied by the Fermi function and a background. Both peaks are Voigt profiles (convolution of a Lorentz profile and a Gaussian profile) with the same shape and size. They are located at equal distances from the Fermi level. The fitting function is the following:\\
$I(\epsilon)= I_0 + \big(I_1 + V(\epsilon - E_F -\Delta, A, W, S) + V(\epsilon - E_F +\Delta, A, W, S)\big) F(\epsilon,E_F,T)$,\\
where $F(\epsilon,E_F,T)= \big(1+exp\frac{\epsilon - E_F}{kT}\big)^{-1}$ is the Fermi function and $V(x, A, W, S)$ is a Voigt profile. Here $\epsilon$ is binding energy; $A$, $W$ and $S$ are numbers which represent the area, width, and ratio of Lorentz and Gaussian components of Voigt profile; $E_F$ is the Fermi level position; $\Delta$ is SC gap size; $T$ is temperature; $k$ is Boltzmann constant. The term $I_0 + I_1 F(\epsilon,E_F,T)$ represents a background. For the fiting $I_0$, $I_1$,  $E_F$, $\Delta$, $A$, $W$, $S$, $T$ are fit coefficients and $\epsilon$ is an independent variable. During the fitting coefficient $I_1$ was hold on a value which was estimated from part of the spectrum without bands. Changing of this coefficient in a reasonable rage can only make negligible changes in the fitting results. So holding of $I_1$ should not course inaccuracy in the gap size determination, and we can treat data in this way.

{\bf Fig. S1} shows EDCs obtained from $k_F$ for XY holelike band (one which forms big pocket), XY electronlike band (one which forms inner pocket) and XZ holelike band (one which forms small pocket) from spectra on {\bf Fig. 1A}. EDC for XY holelike was obtained from the 6K spectrum and 2 other EDCs was obtained from the 11K spectrum. Results of fitting these EDC on our function are given in Table. S1. The second coherent peak on EDC obtained from XZ holelike band is more distant from FL and because of this is more suppressed and appears as a shoulder. Nevertheless in this case the gap still can be extracted from the fitting procedure.

There is one more peak on EDC taken throw XY electronlike band on 6.5 meV. This peak originated from the XZ  electronlike band (one which forms outer pocket).

\begin{table}[h]
    \centering
    \begin{tabular}{l|c}
        Band & Gap size\\ \hline
        XY holelike     & 2.30$\pm$0.07 meV \\ 
        XY electronlike & 3.57$\pm$0.03 meV \\
        XZ holelike     & 5.41$\pm$0.10 meV \\
    \end{tabular}
    \caption*{{\bf Table. S1} Gap}
    \label{tab:one}
\end{table}

\begin{figure}[b]
	\centering
    \includegraphics[width=0.30\linewidth]{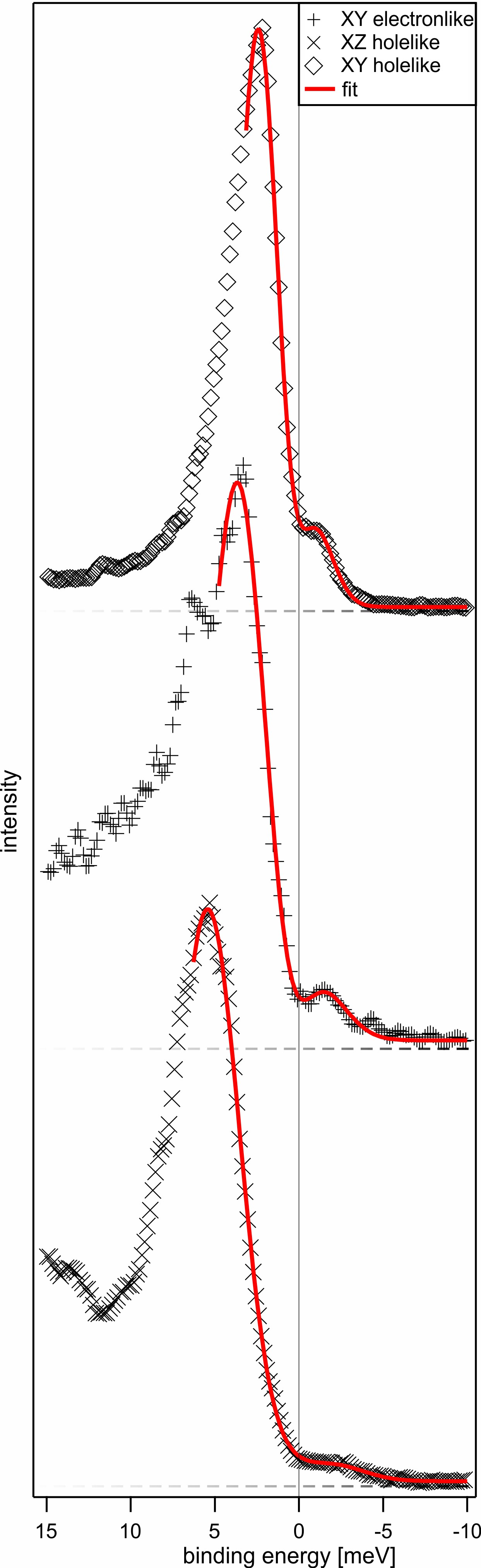}
	\caption*{{\bf Fig. S1} Gap fitting.}
	\label{fig:s1}
\end{figure}

\clearpage

\subsection*{III. k$_z$ dispersion}
{\bf Fig. S2} shows k$_z$-map which allows to determine the h$\nu$ corresponding to G- and Z-points of the BZ. $Z$ corresponds to $\sim$ 37eV, $\Gamma$ corresponds to $\sim$ 26eV and next $Z$ is at $\sim$ 18eV or a little bit lower.

\begin{figure*}[b!]
	\centering
    \includegraphics[width=0.80\linewidth]{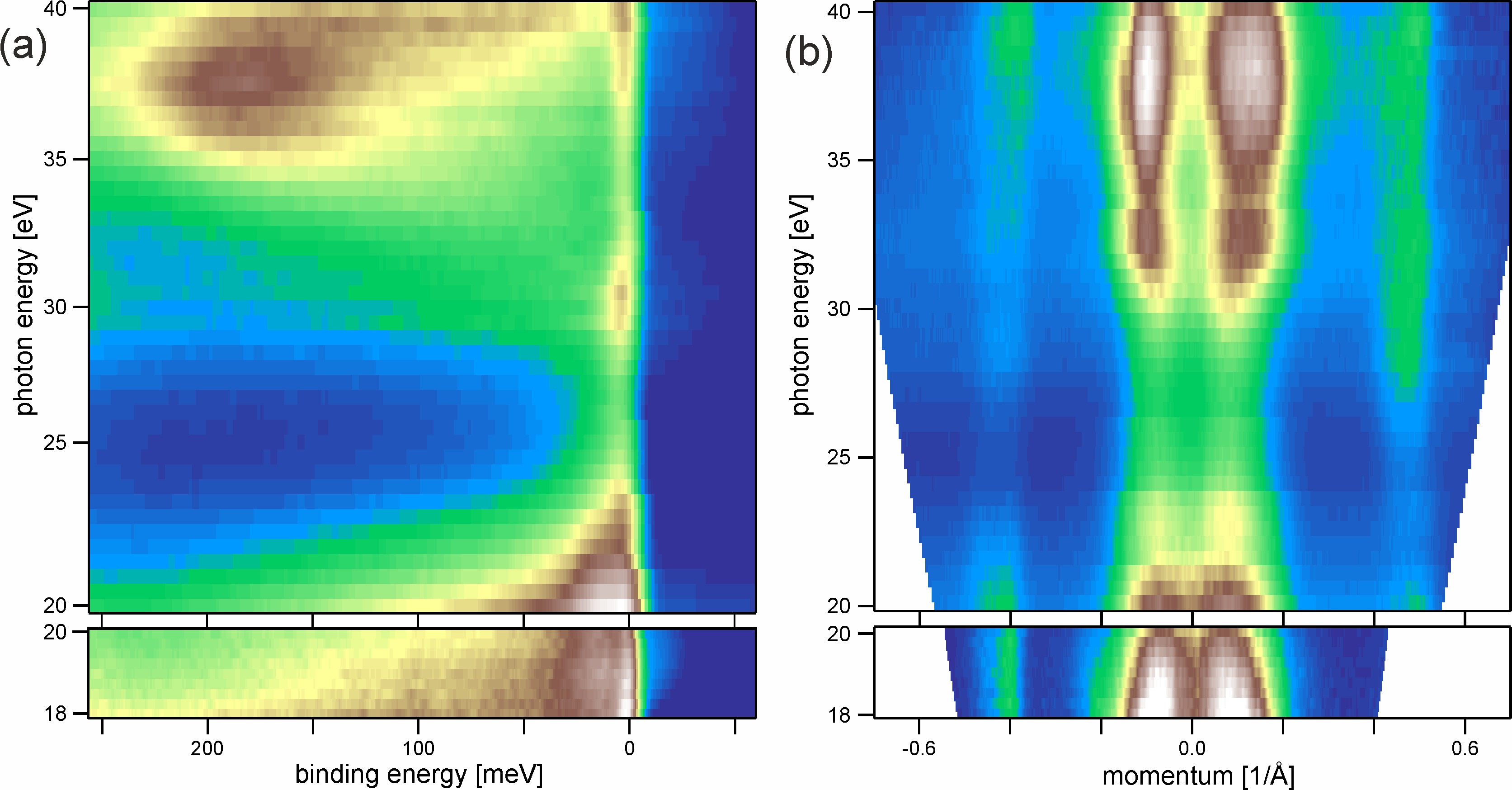}
	\caption*{{\bf Fig. S2} (A) a set of EDC which were obtained through center of the hole-like dispersion ($k_y$=0) for different photon energy. (B) $k_z$-map. }
	\label{fig:s2}
\end{figure*}

\end{document}